\documentclass[11pt]{article}

\usepackage{geometry}
\usepackage{xcolor}
\usepackage{relsize}  % Needed to define Key4HEP name macro
\usepackage{xspace}
\usepackage[T1]{fontenc}
\usepackage[utf8x]{inputenc}
\usepackage{microtype}
\usepackage[backend=biber,style=numeric-comp,sorting=none,giveninits=true]{biblatex}
\usepackage{hyperref}
\usepackage{authblk}
\usepackage{enumitem}

\renewbibmacro{in:}{}

\title{\vspace*{3cm}
  {\bf To build or not to build: FCC}\\
 {\Large  personal input to the 2026 update of the European Strategy for Particle Physics}  }

\date{March, 2025}
\author{Aleksander Filip Żarnecki}
\affil{\vspace*{-3mm}filip.zarnecki@fuw.edu.pl}
\affil{Faculty of Physics, University of Warsaw, Poland}

\newcommand{\ee}{e$^+$e$^-$\xspace}

\newcommand{\keyhep}{{K{\smaller[2]{EY}}4{\smaller[2]{HEP}}}\xspace}

\begin{document}

\maketitle
\thispagestyle{empty}

\vspace*{3cm}
\begin{abstract}
  This is my individual contribution to the discussions ongoing on the update of the European Strategy for Particle Physics. After participating in many discussions and contributing to a number of documents being prepared for the ESPP submission, I came to the conclusion that these ``official'' documents do not give the whole picture. I wanted to express my concerns about the strategy update process and the possible choice of FCC as the next flagship project at CERN. I do hope my short contribution is found relevant by those interested in the strategy update process.
\end{abstract}

\setcounter{page}{0}

\newpage

%%%%%%%%%%%%%%%%%%%%%%%%%%%%%%%%%%%%%%%%%%%%%%%%%%%%%%%%%%%%%%%%%%%%%%%%%%%%%%%%%%%%%

\section{Motivation}

Discussions on the European Strategy for Particle Physics (ESPP) are ongoing since over twenty years.
The first Strategy Group was established in June 2004 and, after the community ``Open Symposium'' held in Orsay, the first strategy document was approved by the CERN Council in July 2006. 
I participated in the Orsay symposium, I also took part in the 2013 update of the ESPP as a member of the Preparatory Group (nominated by RECFA), contributing to the organization of the second Open Symposium in Krakow and to the strategy drafting session in Erice.
I was involved in the preparation of the Polish community input submissions to the initial strategy document and the following updates, including the current one.
I participated in numerous discussions on the European strategy at different meetings as well as in many informal ones.
I decided to submit this document to share my observations, but also some concerns and objections to the way the strategy update process is ongoing, with other members of our community.
I do hope they can also be of interest to people from outside our field, who would like to understand better the way the strategy decisions are made.

I first came to the idea of writing this contribution at the  seminar on ``Future Perspectives in High Energy Physics''  organised by the  International Committee for Future Accelerators (ICFA) at DESY, Hamburg, in November 2023. During the main panel discussion, CERN Director General (DG) Fabiola Gianotti stated her reason for considering only the FCC as the next flagship project for CERN: ``this is the only project which matches the size of CERN community''.
I did know that the size of the FCC was one of the main reasons for its wide support at CERN, but I had never expected this to be confirmed so openly, admitting at the same time that the physics case was of secondary importance.

Several projects have been presented as possible candidates for the future Higgs/electroweak/top factory. The most advanced are the two circular collider projects, the \ee Future Circular Collider (FCC-ee) at CERN and the Circular Electron–Positron Collider (CEPC) in China, and the three linear ones: the International Linear Collider (ILC) in Japan, the Compact Linear Collider (CLIC) at CERN and the Cool Copper Collider (C$^3$) in the US. Combined vision of the Linear Collider Facility at CERN has been prepared as an input for the ESPP update \cite{LinearColliderVision:2025hlt}. Circular colliders are limited in the energy reach to about the top pair-production threshold, but their main advantage is the extremely high luminosity at the lowest energy, when running at the Z-pole. Linear colliders on the other hand, allow to extend the energy reach to the TeV scales with polarised beams. 

The physics case for the future \ee Higgs factory is considered in many other submissions. 
I contributed myself to different studies addressing prospects of (linear) \ee colliders, working in this field since over 20 years. 
I am convinced that the Linear Collider Facility at CERN \cite{LinearCollider:2025lya} offers the best physics prospects, but I will not discuss it here. I do hope for many fruitful discussions on the subject at the Open Symposium in Venice. Research opportunities at different facilities are to be studied in detail by the Preparatory Group and described in the strategy ``Briefing Book''. 
I do believe that the strategy discussions should focus on the physics case.
We should try to make best use of the resources which are at our disposal to optimize the expected progress in our understanding of the Universe. Unfortunately, as I will also try to discuss in the following, the current strategy update process is not entirely based on this approach.

\section{Physics case}

Previous update of the ESPP, adopted by the CERN Council in June 2020, confirmed that the ``electron-positron Higgs factory is the highest-priority next collider'' \cite{CERN-ESU-015}. European Committee for Future Accelerators (ECFA) launched a series of workshops to further explore challenges of the Higgs factory project and bring together different communities, foster cooperation across various existing projects. Three working groups were formed to work on the Physics Potential (WG1), Physics Analysis Methods (WG2) and Detectors (WG3).
One of the goals of the study was to extend our understanding of the physics potential of the \ee machines, which had already been studied for the selected topics in the past, towards the unexplored areas. 

In 2023, after long discussions within WG1, a list of so called focus topics was defined, eventually published on arXiv in January 2024 \cite{deBlas:2024bmz}.
Defined were high priority targets, where further studies were needed for full understanding of the physics potential of an \ee Higgs/Top/Electroweak factory. 
These topics were also to act as ``a vehicle for new engagement and collaboration'', extending the community interested in and working on the physics case for the future \ee machine in general.
This was quite an ambitious plan, as full simulation studies based on the \keyhep framework were strongly encouraged.  
Considered studies were expected to take up to two years and the final submission of the ECFA study report was planned in December 2025, which looked as a ``reasonable assumption for the input date for the strategy process'' \cite{ecfa2023:kj}.

In March 2024, the CERN Council launched the process for the third update of the Strategy. It is clear that Council members had to be informed about the timeline of the ECFA study. Nevertheless, the adopted schedule for the strategy update was much tighter than that assumed in the ECFA study timeline, shortened by almost a year.  
As a result, many research which were planned within the study, full simulation studies in particular, could not be completed or were completed only partially. 

Why was the CERN Council so determined to proceed with the fast update of the ESPP, not waiting for the final result of the ECFA study (launched in response to the recommendations of the previous strategy update)? 
This decision confirmed, in my opinion, that the physics case for the future \ee machine was not considered very relevant for the strategy update.
The other reason could be that the topics selected in the ECFA study were clearly pointing to the importance of the high energy reach. While one focus topic could only be addressed at the Z pole, all remaining ones could clearly benefit from running at energies of 500\,GeV and above (see Tab. 1 in \cite{deBlas:2024bmz}).
As the CERN management had already decided to support FCC, results pointing to the necessity of high energy \ee running were rather to be avoided. 
Finally, one should note that any delay in the schedule would result in the current CERN DG not being able to oversee the final strategy drafting session currently planned for December 2025.  

\section{Social aspects}

CERN is the largest HEP research centre in the world. 
Since over 70 years it supports nuclear and particle physics research in (currently) 24 CERN Member States.  
Most of the research activities ongoing in Europe are connected with CERN and majority of experimental particle physics community is currently involved in the running experiments at the LHC.
They profit to a large extent from stable, long-term funding and clear research perspectives for the coming 15 years. 
They are in a privileged situation compared to those involved in smaller, short time-scale experiments, not to mention researchers in other branches of science, where the usual timescale of the project is of the order of months or single years. 
But the LHC community would still like to extend this social arrangement to the next decades, to the next flagship project at CERN, so not only they but also their students were secured for the whole academic careers.
To ensure this, the next flagship project at CERN needs to be huge and expensive, and the concept of FCC, covering both the \ee collider (FCC-ee) as well as the subsequent hadron machine (FCC-hh), perfectly fits these demands.
That is, in my opinion,  why so many people at CERN share the point of view presented by Fabiola Gianotti at the ICFA seminar in Hamburg.
As I have heard from the prominent LHC physicist myself, linear colliders are just too small to be suitable for the CERN community.

For me, the size of the FCC project, the fact that it would involve most of the community at CERN, is exactly an argument against the project.
Already now we do observe than not only CERN but also our field as a whole is significantly biased towards the LHC physics. 
At the last International Conference on High Energy Physics (ICHEP'2024) in Prague, nine of the plenary talks were devoted to the LHC experimental results only. 
This is probably not surprising, as the LHC is the only running energy frontier collider in the world and most of the community is involved in its experiments.
But still, the situation has its drawbacks. 
We are flooded with hundreds of papers from the LHC experiments, but most of them hardly contribute to our understanding of the Universe. There are twice as much ``search'' papers (with negative results) than the ``measurement'' ones.\footnote{Looking at the titles of the published ATLAS and CMS papers in arXiv.}
Each new measurement is presented as a great success and CERN does a great job in PR. But one has to realise that many of these measurements, even those covered by CERN press releases, just confirm what we already knew before \cite{CERN:2024pr,CMS:2024aps}.
Despite the huge investment and the efforts of the thousands of people, the LHC has not met the expectations. 
Its experiments has managed to confirm the existence of the Higgs boson and confirm its basic properties but no ``new physics'' has been observed so far. The resulting progress in our understanding of the particle interactions at the most fundamental level is, in my opinion, marginal. 

And this situation may repeat itself in about 50 years, when FCC-hh is built. When I asked another prominent physicist about his main motivation for FCC-hh the answer was: ``hadron colliders have always been the discovery machines''. But this argument is no longer valid! All major discoveries at hadron colliders (W$^\pm$ and Z bosons, top quark, Higgs boson) were precisely guided by the predictions of the Standard Model. We knew that there are new states to be discovered, we knew their properties and how to search them. We have no such predictions at the moment, not even a hint about the possible mass scale involved, no guarantee that there is anything beyond TeV scale awaiting discovery.
So the main motivation to push towards the FCC-hh, as it looks to me, is to secure the follow-up for those enjoying their work on proton-proton interactions at the LHC and willing to continue along the same lines.  

At least in Poland, it is increasingly difficult to get any funding for high energy physics research not related to CERN.
For the FCC to be built, not only most of the resources available at CERN will need to be involved, but also most of the resources allocated for HEP in the Member States.
There is a high risk that the progress in other areas of research will be hindered significantly.
CERN, which should be the centre of excellence for all branches of nuclear and particle physics research, equally supporting a wide range of ambitious experimental projects, will be reduced to the role of the FCC host laboratory.
Focusing on the project based on the 50-year old accelerator concept (to be run in another 50 years) also looks like the best way to loose the leadership in the field.

\section{National inputs}

For the expected inputs to the ESPP update from the HEP communities in the Member State, ECFA has prepared dedicated guidelines.
A set of detailed questions was given, to be considered in the national documents.  
The questions were formulated in such a way, as the answer to the first one, ``Which is the preferred next major/flagship collider project for CERN?'' could only be one, namely FCC.
It looked more as an invitation to cast a vote than to give a wider input to the strategy shaping discussions.  

Over the last few months, I participated in many discussions on the Polish community input to the ESPP update \cite{PL4ESU}. 
The final document submitted recently clearly states that the community ``give preference and declare willingness to actively engage and participate in every aspect of the FCC project''. 
This seems to be the point of view of the large part of the Polish community and I do not question it. 
However, I would like to report on the arguments which were given for the FCC support. 
While the importance of the precision electroweak measurements at the Z-pole was clearly mentioned, the main arguments followed very different lines:
\begin{itemize}[noitemsep,topsep=0em]
\item CERN has already decided. We depend on CERN too much to question it, so we must support CERN's decision;
\item We have made a substantial contribution to building and running the LHC and its experiments. We do have all necessary know-how, so we can easily contribute to the next hadron machine at CERN (we could skip FCC-ee and go directly to FCC-hh);
\item A lot of work has been put into FCC Feasibility Study, preparations are well advanced including negotiations with local municipalities, this can not be wasted. Any new project would need to start from the begining and it would take much more time;
\item If CERN switched to a different project, it would loose its credibility. This could endanger the support currently expressed by the Member States.
\end{itemize}
I do not agree with any of these arguments. If we consider the first one, the whole ESPP update process looks like a PR move: CERN makes the decision and then asks the communities in the Member States to give their unanimous support, to strengthen the CERN's position in the negotiations with the Member State governments. 
We also heard at the last ECFA study workshop that ``there is no room for disagreements after we converge'' (on the choice for the next collider at CERN) \cite{ecfa2024:ps}.
This statement sounded almost like a threat to me and I do think we should not agree with such an approach. There can be no progress in science, if there is no freedom in expressing our opinions. 
As for the politicians, they are well used to changes in the strategy plans after elections. We could only gain credibility, if we openly present the motivation for the change in our plans.
And there is much more to be lost than a short term credibility. 
If CERN devotes all the available resources to FCC, it is my strong believe that it is likely to loose, sooner or later, the leadership in the field and the position of the world's best centre of excellence in particle physic.

\section{Conclusions}

I do think many people realise the problems I described above. 
But only few of them decide to speak openly and express their concerns.
I do hope the discussion on the update of the European strategy will be open, 
based primarily on the physics case, which is to be studied in detail by the Preparatory Group. 
And I do believe the choice of the future flagship project for CERN should be based on the physics merits and the necessity to ensure that the diverse research options remain open for the community. 

Any future particle physics project at the energy frontier will require huge investments. 
We try to convince our governments and ordinary tax-payers that our plans are worth it, that we can make the best use of the resources and contribute most to the general growth of the society (in terms of the scientific knowledge, technology development and education).
We try to convince them that the investment in high energy physics research is as needed as investments in medical research or environmental studies, which can have much more direct impact on the people (not to mention other fields of fundamental research).

My main point is that, as researchers and academic teachers we are bound to be honest. 
We must state clearly what our reasons for supporting given project are.
People funding our future have right to know what they are paying for. If CERN decides to build FCC they will pay not only for the possible progress in our understanding of the Universe, but also for securing stable positions and long term funding for thousands of physicists and engineers, basically independent of the actual results of their work.
If approved, FCC will secure bright future for CERN as an enterprise for at least another 50 years. I am not convinced it will secure any future for the particle physics.

\printbibliography %  [heading=none]

\end{document}